%% file: main.tex
\documentclass[sigconf,authorversion]{acmart}

\usepackage{pbox}
\usepackage{xcolor}
\usepackage[utf8]{inputenc}
\usepackage[show]{chato-notes}
\usepackage{tabu}
\usepackage{caption}
\usepackage{subcaption}
\usepackage[para]{footmisc}
\usepackage[toc,page]{appendix}

\usepackage{listings}
\usepackage{listingsLanguages}
\usepackage{mathtools}
\usepackage{booktabs}
\usepackage{multicol}
\usepackage{multirow}
\usepackage{adjustbox}
\usepackage{hyperref}

\makeatletter
\newcommand\rowtag[2]{#1\def\@currentlabel{#1}\label{#2}}
\makeatother

\usepackage{xcolor,colortbl}

\usepackage{amsthm}

\usepackage{amsmath}

\newcommand{\crc}[1]{\textcolor{black}{#1}}

\newcommand{\craig}[1]{\textcolor{black}{#1}}
\newcommand{\nic}[1]{\textcolor{black}{#1}}
\newcommand{\cmC}[1]{\textcolor{black}{#1}}
\newcommand{\craigC}[1]{\textcolor{black}{#1}}
\newcommand{\nicC}[1]{\textcolor{black}{#1}}

\newcommand{\craigf}[1]{\textcolor{black}{#1}}

\newcommand{\taat}{\textsf{TAAT}}
\newcommand{\quit}{\textsf{Quit}}

\usepackage{tablefootnote}
\usepackage{array}
\newcolumntype{L}{>{\arraybackslash}m{6cm}}
\newcolumntype{M}{>{\arraybackslash}m{4cm}}
\title{Query Embedding Pruning for Dense Retrieval}
\author{Nicola Tonellotto}
\affiliation{%
  \institution{University of Pisa, Italy}
} \email{nicola.tonellotto@unipi.it}

\author{Craig Macdonald}
\affiliation{%
  \institution{University of Glasgow, UK}
} \email{craig.macdonald@glasgow.ac.uk}

\usepackage{marginnote}

\newcommand{\pageenlarge}[1]{\enlargethispage{#1\baselineskip}}

\copyrightyear{2021}
\acmYear{2021}
\setcopyright{acmcopyright}\acmConference[CIKM '21]{Proceedings of the 30th ACM International Conference on Information and Knowledge Management}{November 1--5, 2021}{Virtual Event, QLD, Australia}
\acmBooktitle{Proceedings of the 30th ACM International Conference on Information and Knowledge Management (CIKM '21), November 1--5, 2021, Virtual Event, QLD, Australia}
\acmPrice{15.00}
\acmDOI{10.1145/3459637.3482162}
\acmISBN{978-1-4503-8446-9/21/11}

\begin{document}
\fancyhead{}
%\settopmatter{printacmref=false} % Removes citation information below abstract
%\renewcommand\footnotetextcopyrightpermission[1]{} % removes footnote with conference information in first column
%\pagestyle{plain} % removes running headers

%\pagenumbering{gobble}

\begin{abstract}

%\inote{(Nic would remove that) Over the last few years, contextualised language models such as BERT have demonstrated significant benefits to search effectiveness; this is because the embeddings for the query and document tokens are able to distinguish between different contextual meanings of the tokens, and leading to a better estimation of relevance between queries and documents.} 

\looseness -1 Recent advances in {\em dense retrieval} techniques have offered the promise of being able not just to re-rank documents using \nic{contextualised language models such as BERT}, but also to use such models to identify documents from the collection in the first place. %, thereby demonstrating the potential to supplant the classical inverted index data structure.
However, when using dense retrieval approaches that use multiple embedded representations for each query, a large number of documents can be retrieved for each query, hindering the efficiency of the method. Hence, this work is the first to consider efficiency improvements in the context of a dense retrieval approach (namely ColBERT), by {\em pruning} query term embeddings that are {\em estimated} not to be useful for retrieving relevant documents. Our proposed query embeddings pruning reduces the cost of the dense retrieval operation, as well as reducing the number of documents that are retrieved and hence require to be fully scored.
\nicC{Experiments conducted on the MSMARCO passage ranking corpus demonstrate that, when reducing the number of query embeddings used from 32 to 3 based on the collection frequency of the corresponding tokens, query embedding pruning results in no statistically significant differences in effectiveness, while reducing the number of documents retrieved by 70\%. In terms of mean response time for the end-to-end to end system, this results in a $2.65\times$ speedup.}\vspace{-0.34\baselineskip}
%\craig{Experiments conducted on the MSMARCO passage ranking corpus demonstrate that, for example, when reducing the number of query embeddings used from 32 to 6, our \eaat\ approach results in no drop in effectiveness, while reducing the number of documents retrieved by 77\%, thereby speeding up response times by upto 2.5 times.}
%our approach can correctly identify a single representative query term embedding that results in no statistically significant loss in effectiveness for NDCG@10 or MRR, 
%while reducing by 92\% the number of documents retrieved compared to ColBERT's suggested dense retrieval configuration.}\inote{check}
\end{abstract}



\vspace{-0.34\baselineskip}

\maketitle
%\todo{update/remove ACM headers}

%\inote{Place the TAAT stuff in introduction.}
%\inote{Think about the efficiency claims: run on trec2019 for p=3 and p=32 to report the savin}

\vspace{-0.34\baselineskip}\input{introduction}
\input{dense}
\input{eaat}
\input{experiments}
\vspace{-.75\baselineskip}
\input{conclusions}

\newcounter{BalanceAtReference}
\setcounter{BalanceAtReference}{10}
\newcounter{ReferenceIndexForBalancing}

\makeatletter

% Disable acmart's automatic invocation of \balance from \AtEndDocument,
% which is usually too late.
\global\@ACM@balancefalse

% Invoke command when the \bibitem reaches the specified value
\def\@balancelastpageonce{%
  \ifnum\value{ReferenceIndexForBalancing}=\value{BalanceAtReference}
    \newpage
  \else
    \relax
  \fi
  \stepcounter{ReferenceIndexForBalancing}
}
\pretocmd{\bibitem}{\@balancelastpageonce}
  {} % on success
  {\@latex@error{Patching \bibitem failed}{\@ehd}}

\makeatother

\bibliographystyle{ACM-Reference-Format}
\bibliography{bib} 
\balance

\end{document}

%% file: introduction.tex
\section{Introduction}\label{sec:intro}\pageenlarge{1}

\looseness -1 Pretrained contextualised language models such as BERT~\cite{bert} are able to successfully exploit general language features in order to capture the contextual semantic signals allowing to better estimate the relevance of documents w.r.t. a given query, leading to effective search ranking improvements when re-ranking the documents obtained from a classical inverted index~\cite{lin2020pretrained}. 
Recently, representation-focused models have gained attention due to their ability to capture semantic information and because they allow to pre-compute document representations at indexing time, greatly reducing the query processing times~\cite{epic,prettr,khattab2020colbert,snrm}. Representation-focused models aim at learning a function mapping a sequence of tokens, e.g., a query or a document, into one or more real-valued vectors called \textit{embeddings}. These embeddings are then combined to compute a similarity score between the query and the document.
Inspired by distributional word embeddings~\cite{mikolov}, many works have adopted some pooling technique such as max- and mean-pooling to generate a \textit{single representation} for a sequence of terms. However, the use of a single representation in effect compresses all possible semantic facets of the given text into a single vector. More recently, \textit{multiple representations}, composed by a list of embeddings, one per term in the sequence, have been investigated~\cite{khattab2020colbert,polyenc,luan2020}. In this case, there is a similarity score between every query term embedding and every document term embedding, and the pooling is performed when all similarity scores have been computed. % Several works have shown that multiple representations provide better effectiveness compared to single representations~\cite{khattab2020colbert,polyenc,luan2020}. 
Most neural ranking approaches have been used to by re-rank the documents identified by a classical inverted index using relevance models such as BM25, in a multi-stage ranking architecture~\cite{lin2019recant,Matveeva2006HighAR}. However, \nicC{lexical matching models relying solely on an inverted index may not identify the contextually related candidate documents that would have been highly scored by an effective neural ranking model}. Instead, by utilising documents encoded as vectors at indexing time and queries encoded as vectors at query processing time, dense retrieval approaches~\cite{xiong2020approximate,khattab2020colbert} are of growing interest. In dense retrieval, the top-ranked documents for a given query are computed by identifying the most similar document embeddings to the given query embeddings, employing a nearest neighbour search procedure. Nearest neighbour search with single representations has been shown to be efficient, but less effective than multiple representations~\cite{lin2020pretrained}. On the other hand, when multiple representations are exploited, as pioneered by Khattab and Zaharia~\cite{khattab2020colbert}, a multi-stage dense retrieval approach can be executed, where the first stage conducts an \textit{approximate} but highly efficient nearest neighbour search, retrieving documents to be exactly scored by the second stage.%a candidate set of

\pageenlarge{0} However, using multiple representations for dense retrieval results in a first-stage that retrieves many documents. These documents are then re-ranked in the second stage, and the time spent in re-ranking is directly proportional to the number of documents retrieved by the first stage. Hence, the time taken to score all retrieved documents can be expensive.
%If the encoded document representations are in fast storage such as RAM, the scoring conducted in the second-stage can be quite fast. However, particularly if slower storage is used, the time taken to score all retrieved documents can be expensive.
\nicC{Instead, this paper proposes the adaptation of dynamic pruning strategies for dense retrieval. 
The retrieval of the most similar documents through ANN can be seen as a term-at-a-time (\taat) set retrieval. \taat\ retrieval for classical best match weighting models (such as BM25) involves the processing of all the documents in which a specific query term appears, to compute a query-document score contribution stored in each document's score accumulator~\cite{fntir2018}. Dynamic pruning strategies have been proposed to reduce the number of accumulators being created or updated. In the case of \taat\ \quit\ dynamic pruning strategies, query processing terminates after a certain number of query terms have been processed and the top $k$ documents are selected from the documents processed thus far. In our \craigf{query embedding pruning approach, we are inspired by \taat\ \quit, by proposing to estimate which multiple query embeddings} are useful for dense retrieval. Indeed, by pruning out less useful query embeddings, we can conduct faster approximate nearest neighbour search, reduce the number of documents that are retrieved by the first stage dense retrieval, and \craigf{obtain a} faster second-stage scoring.}

In summary, this work contributes (i) a first examination of the usefulness of different query embeddings in multiple representation dense retrieval, and (ii) the novel proposition of dynamic pruning of query embeddings for dense retrieval. 
\nicC{In particular, our experiments conducted on the MSMARCO passage ranking corpus demonstrate that, for example, when reducing the number of query embeddings used from 32 to 3, our query embedding pruning approach results in no statistically significant differences in effectiveness, while reducing the number of documents retrieved by 70\%.}
%\nicC{The structure of the remainder of this paper is as follows: Section~\ref{sec:dense} defines the dense retrieval task, and discusses limitations of one existing approach, namely ColBERT. In Section~\ref{sec:eaat} we propose our query embedding pruning approach. Experiments follow in Section~\ref{sec:experiments}.  We summarise our findings and provide directions for future work in Section~\ref{sec:conclusions}}.

%% file: dense.tex
\pageenlarge{3} \section{Dense Retrieval}\label{sec:dense}
\setlength{\belowdisplayskip}{2pt} \setlength{\belowdisplayshortskip}{2pt}
\setlength{\abovedisplayskip}{2pt} \setlength{\abovedisplayshortskip}{2pt}

We assume that queries and documents are sequences of terms from a given vocabulary $V$. Any term is represented by a real-valued vector of dimension $d$, called an embedding. More formally, let $f_Q: V^{n} \to \mathbb{R}^{n \times d}$ be a learned function mapping a given query term $t_i$ in a query of $n$ terms to the query embedding \nic{$\phi_i$, i.e., $\{\phi_1, \ldots, \phi_n\} = f_Q(t_1,\ldots,t_n)$}. Similarly, let $f_D: V^{n} \to \mathbb{R}^{n \times d}$ be a (potentially different) learned function mapping a given document term $t_j$ in a document of $n$ terms to the document embedding \nic{$\psi_j$, i.e., $\{\psi_1, \ldots, \psi_n\} = f_D(t_1,\ldots,t_n)$}.
%A given query $q$ is typically\inote{colbert only??} padded to a fixed token length $|\hat{q}|$ such as 32 to provide a query expansion mechanism for term tokens similar to the tokens appearing in the original query. 
Hence, a query $q$ composed by $|q|$ tokens is represented by $|q|$ query embeddings $\{\phi_1,\ldots, \phi_{|q|}\}$. Analogously, a given document $d$ composed by $|d|$ tokens is represented by $|d|$ document embeddings $\{\psi_1, \ldots, \psi_{|d|}\}$.
Given two embeddings, their similarity is computed by the dot product. Hence, for a query $q$ and a document $d$, their final similarity score $s(q,d)$ is obtained by summing up the maximum similarity between the query token embeddings and document token embeddings:
\begin{equation}\label{eq:maxsim}
    s(q,d) = \sum_{i=1}^{|q|}\max_{j=1,\ldots,|d|} \phi_i^T \psi_j
\end{equation}
The document embeddings from all documents in the collection are pre-computed through the application of \craig{the $f_D$} learned function and stored into an index data structure for vectors supporting similarity searches. This can
%, i.e., a dense retrieval component able to 
identify the closest vectors to a given input vector leveraging with cosine or dot product vector comparisons. Query token embeddings are computed at runtime leveraging the $f_Q$ learned function; queries may also be augmented with additional \emph{masked tokens} to provide  ``{\em a soft, differentiable mechanism for learning to expand queries with new terms or to re-weigh existing terms based on their importance for matching the query}''~\cite{khattab2020colbert}.\footnote{In current practice~\cite{khattab2020colbert}, queries are augmented up to $32$ query token embeddings.}

\looseness -1 In order to reduce the time required to compute the similarities between query and document embeddings using Eq.~\eqref{eq:maxsim}, it is possible to shift from an \textit{exact} to an \textit{approximate} nearest neighbour (ANN) search. With ANN, the document embeddings are stored in a quantised form, suitable for fast searching. However, the approximate similarity scores between these compressed embeddings \nicC{are inaccurate, and hence are not used for computing the final top documents.} %insufficient for computing the final top documents\inote{CM: NOW WRONG}. 
Indeed, ANN search computes, for each query embedding $\phi_i$, the set $\Psi(\phi_i, k')$ of the $k'$ document embeddings most similar to $\phi_i$ according to some approximate distance; then, these document embeddings are mapped back to their documents $D_i(k')$:
\begin{equation}\label{eq:faiss1}
    D_i(k') = \{d \in D\;:\; f_D(d) \cap \Psi(\phi_i, k') \neq \emptyset\}
\end{equation}
and finally the union $D(k')$ of these sets is returned:
\begin{equation}\label{eq:faiss2}
    D(k') = \bigcup_{i=1}^{|q|} D_i(k')
\end{equation}
\looseness -1 Once the approximate nearest documents $D(k')$ have been identified, they are exploited to compute the final list of top $k$ documents to be returned. To this end, the set of documents $D(k')$ is re-ranked using the \nic{query embeddings} and the documents' multiple embeddings to produce exact scores that determine the final ranking. 

\begin{figure}[tb]
\includegraphics[width=.95\linewidth]{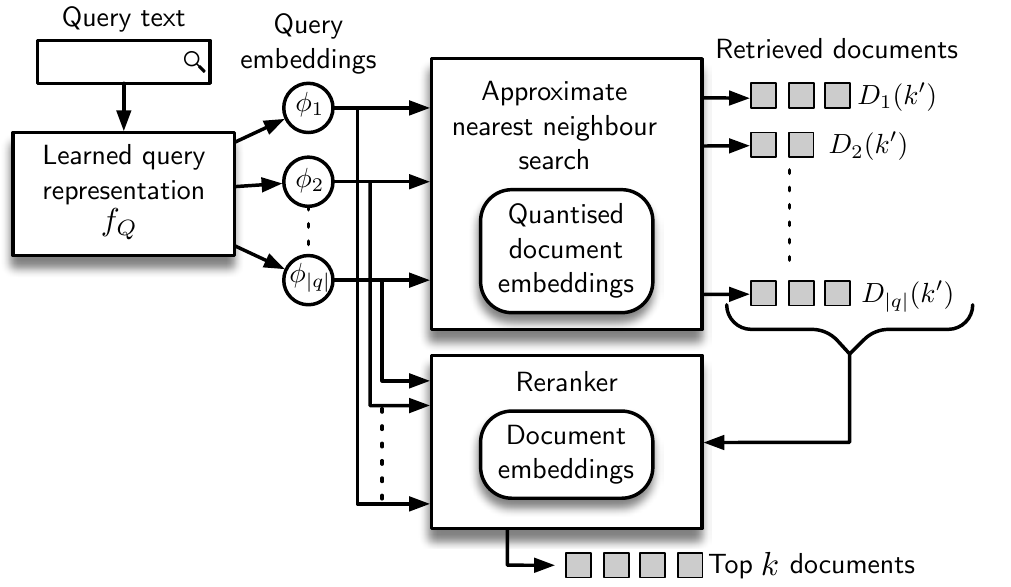}\vspace{-.5\baselineskip}
\caption{Dense retrieval architecture.}\label{fig:architecture}\vspace{-1.5\baselineskip}
\end{figure}

Figure~\ref{fig:architecture} summarises the dense retrieval architecture described above. At the time of writing, ColBERT~\cite{khattab2020colbert} is the only effective dense retrieval system exploiting the multiple representations for queries and documents proposed thus far, exhibiting higher effectiveness than dense retrieval based on single query and document representations such as ANCE~\cite{xiong2020approximate} (see~\cite[Table 27 vs. Table 28]{lin2020pretrained}).

\begin{figure*}[ht!]
\def\figurewidth{\textwidth}
    \centering
    \begin{subfigure}[t]{.24\linewidth}
    \includegraphics[width=\figurewidth]{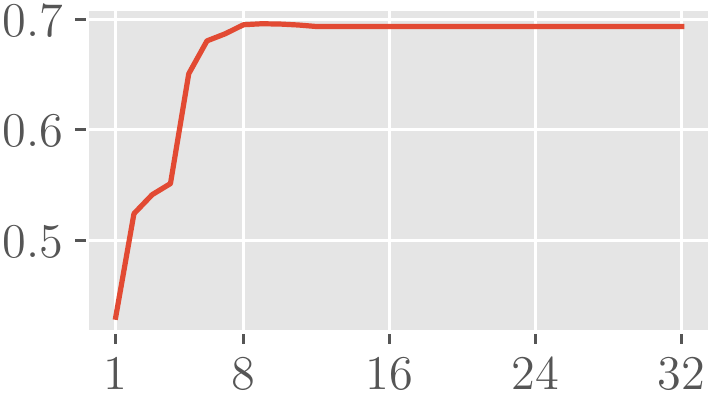}\vspace{-1.5mm}
    \caption{nDCG@10}
    \end{subfigure}
    \begin{subfigure}[t]{.24\linewidth}
    \includegraphics[width=\figurewidth]{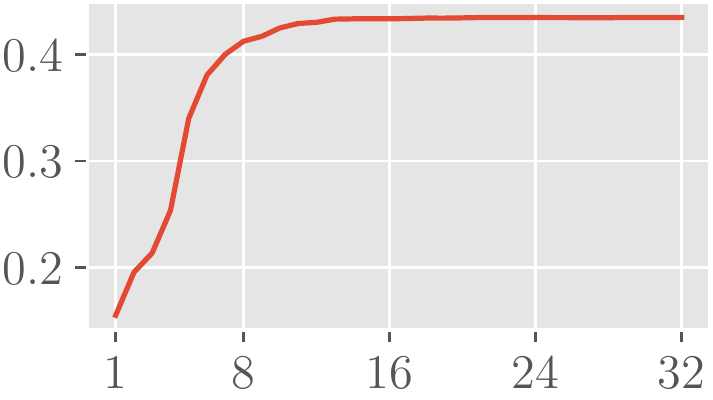}\vspace{-1.5mm}
    \caption{MAP}
    \end{subfigure}
    \begin{subfigure}[t]{.24\linewidth}
    \includegraphics[width=\figurewidth]{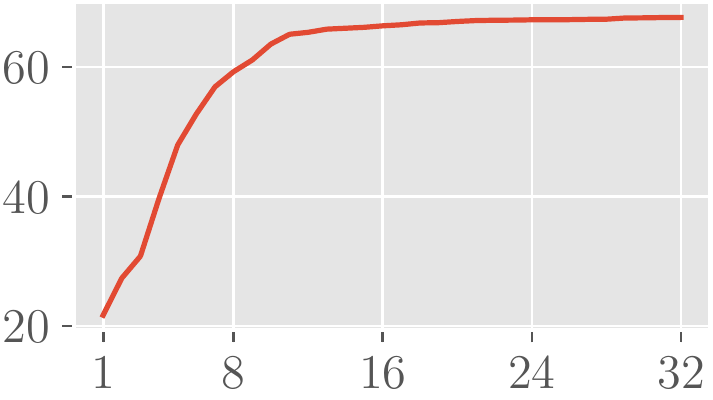}\vspace{-1.5mm}
    \caption{Mean num. rel. docs. retrieved}
    \end{subfigure}
    \begin{subfigure}[t]{.24\linewidth}
    \includegraphics[width=\figurewidth]{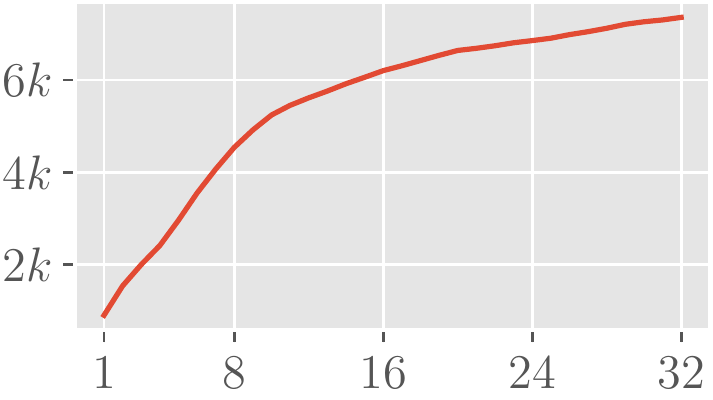}\vspace{-1.5mm}
    \caption{Mean num. docs. retrieved}
    \end{subfigure}\vspace{-.75\baselineskip}
    \caption{TREC 2019 deep learning track effectiveness for various numbers of query embeddings \nic{($k' = 1000$)}.}
    \label{fig:analysis}\vspace{-0.5\baselineskip}
\end{figure*}

\pageenlarge{2} To illustrate the challenges of using the multiple representation-focused dense retrieval, we analyse the performance of the ColBERT end-to-end dense retrieval system by varying the number of \nic{processed} query embeddings, summarising the main limitations identified by our analysis. All experiments are conducted on the MSMARCO Passage Ranking corpus; effectiveness results are reported on the 43 TREC 2019 Deep Learning track queries -- our experimental setup is detailed later in Section~\ref{sec:experiments}.

\looseness -1 ColBERT uses 32 representations for each query. Assuming that we order them in their order of occurrence, \craig{we can} measure the resulting effectiveness if only a subset of these embedding were used. Figure~\ref{fig:analysis} demonstrates the resulting effectiveness measures in term of normalised discounted cumulative gain (nDCG@10), mean average precision (MAP), mean number of relevant documents retrieved, and mean number of documents retrieved. In each graph, the x-axis represents the number of query embeddings used, \craigf{based on their order of occurrence: CLS, query tokens, masked tokens}.

\looseness -1 From Figure~\ref{fig:analysis}, it can be seen that effectiveness across all three measures (nDCG@10, MAP and number of relevant retrieved) rises as more query embeddings are deployed. \crc{With only one query embedding (i.e. CLS), effectiveness on all measures is low\footnote{\crc{This is probably as the CLS embedding encodes less information in ColBERT than in a single-representation dense retrieval approach such as ANCE}}.}  Effectiveness in terms of nDCG@10 and MAP rises quickly before stabilising, with 6-8 embeddings appearing to be sufficient for nDCG@10 (Figure~\ref{fig:analysis}(a)), and 10-12 being sufficient for MAP (Figure~\ref{fig:analysis}(b)). In contrast, the number of relevant documents continues to rise, but more slowly after 12 embeddings (Figure~\ref{fig:analysis}(c)). This increased recall comes at a cost of retrieving more documents, with an average \nic{$1000$} more documents being retrieved when moving from 15 embeddings to 32 (Figure~\ref{fig:analysis}(d)). Indeed, while the \nicC{recall} of the retrieved document set is increased, it is apparent that these additional relevant documents are \nicC{not} promoted to the top ranks of the final ranking, and hence there is no benefit to the (top heavy) nDCG@10 or MAP measures.

In short, our analysis shows that not all query embeddings are needed for effective retrieval -- indeed, this implies that \craigf{some} query embeddings can be ignored without negatively impacting the effectiveness of the whole system. Moreover, many documents are retrieved by each query embedding -- the ramification is that all encoded document embeddings must be stored in memory, as many documents must be scored by the exact ranker to obtain effective results. Finally, less new documents are retrieved by later query (masked) embeddings, as these embeddings are \nicC{reduntant w.r.t.} to earlier embeddings. 
In order to address the aforementioned limitations, \textit{we aim to use less query embeddings for effective and efficient dense retrieval}. Indeed, we desire high effectiveness with less query embeddings. In the next section we discuss how dynamic pruning can be exploited in dense retrieval to increase efficiency without negatively impacting effectiveness.

%% file: eaat.tex
\pageenlarge{2}\section{Query Embedding Pruning}\label{sec:eaat}

%In dense retrieval, instead of query terms \craig{there are} query tokens and their corresponding query embeddings. 
\looseness -1 As we have shown in Section~\ref{sec:dense}, the query embeddings do not contribute equally to the final effectiveness of the retrieved document set. %However, in contrast to \taat, in ANN search we do not have usable score contributions for retrieved documents\inote{CM: NOW WRONG}, but just binary indicators for a document being retrieved for the following exact re-ranking stage. Nevertheless, 
We \cmC{argue} that not every query embedding will bring useful documents for retrieval, even if each query embedding well represents the context of the query term.
%In contrast with the classical IR query processing, due to the compression and quantisation of the document embeddings, in ANN search, the approximate similarity scores calculated by Equation~\eqref{eq:faiss2} are not suitable for \inote{What}.
Hence, we propose to adapt the \taat\ \quit\ dynamic pruning strategy to Equation~\eqref{eq:faiss2}. \taat\ query processing, as well as its dynamic pruning strategies, orders the query terms by relative importance, e.g., inverse document frequency, such that the rarest query term was executed first. This is motivated by the fact that in best match weighting models the rarest query terms appearing in a document contribute most to the final document score, compared to more common terms. Similarly, we postulate that the \textit{most important} query tokens\footnote{While we use the notion of terms and tokens, these could be wordpieces as identified by the BERT tokeniser.} are more likely to bring relevant documents than non-relevant documents, \nicC{and therefore we propose to prune (remove) the unimportant query embeddings}. Indeed, \citet{10.1007/978-3-030-72240-1_23} noted that exact matches and the more important terms contribute more to the overall ColBERT scores; we argue that these terms are those that should be the focus of the ANN search. 

%Let $\mathcal{P}(q)$ represent a (sub)set of the query embeddings $\{\phi_1,\ldots,\phi_{|q|}\}$. For brevity, we denote $|\mathcal{P}(q)|$ as $p$. the Given the query $q$ and its embeddings, this set denotes, the query embeddings corresponding to the ``most important'' query tokens w.r.t. the relevant documents for the query. 
We propose the following query embedding pruning strategy to compute the results of the ANN search in conjunction with Eq.~\eqref{eq:faiss1}:

\begin{equation}\label{eq:eaat}
    D(k') = \bigcup_{i=1}^{p} D_i(k').
\end{equation}
\looseness -1 According to Eq.~\eqref{eq:eaat}, in query embedding pruning, the ANN search does not compute the set of retrieved documents to be re-ranked over all query embeddings, but it only processes the $p$ most important query embeddings, and computes the $k'$ document embeddings most similar to those only. \nicC{Note that query embeddings are only pruned for the first ANN stage, and are restored for the exact scoring stage.} Our approach ignores \nicC{the} less important query embeddings, \nicC{thus}, as a consequence, lesser query embeddings processed in ANN search will generate lesser documents to be re-ranked using all document embeddings.
The identification of the most important query tokens requires a concept of ordering among query embeddings. A natural way to rank the query embeddings is to order them by ascending order of frequency in the collection of the corresponding query tokens. This is akin to the ordering of query terms in \taat\ by IDF. We denote this ranking of query embeddings as Inverse Collection Frequency (ICF)\footnote{Collection frequency is usually correlated with document frequency. \craigC{Indeed, in our initial experiments we find that ICF and IDF result in almost identical orderings of query terms, and hence only ICF is reported}.}, and postulate that the frequency in the collection of the query token corresponding to a query embedding is inversely proportional to its importance in identifying relevant documents. \nicC{Note that special tokens such as CLS and the masked tokens do not correspond to any document token, hence they are placed after the query embeddings corresponding to actual wordpieces, CLS then masked tokens.}

%% file: experiments.tex
\pageenlarge{2} \section{Experiments}\label{sec:experiments}
\begin{figure*}[ht!]
\def\figurewidth{\textwidth}
    \centering
    \begin{subfigure}[t]{.33\linewidth}
    \includegraphics[width=\figurewidth]{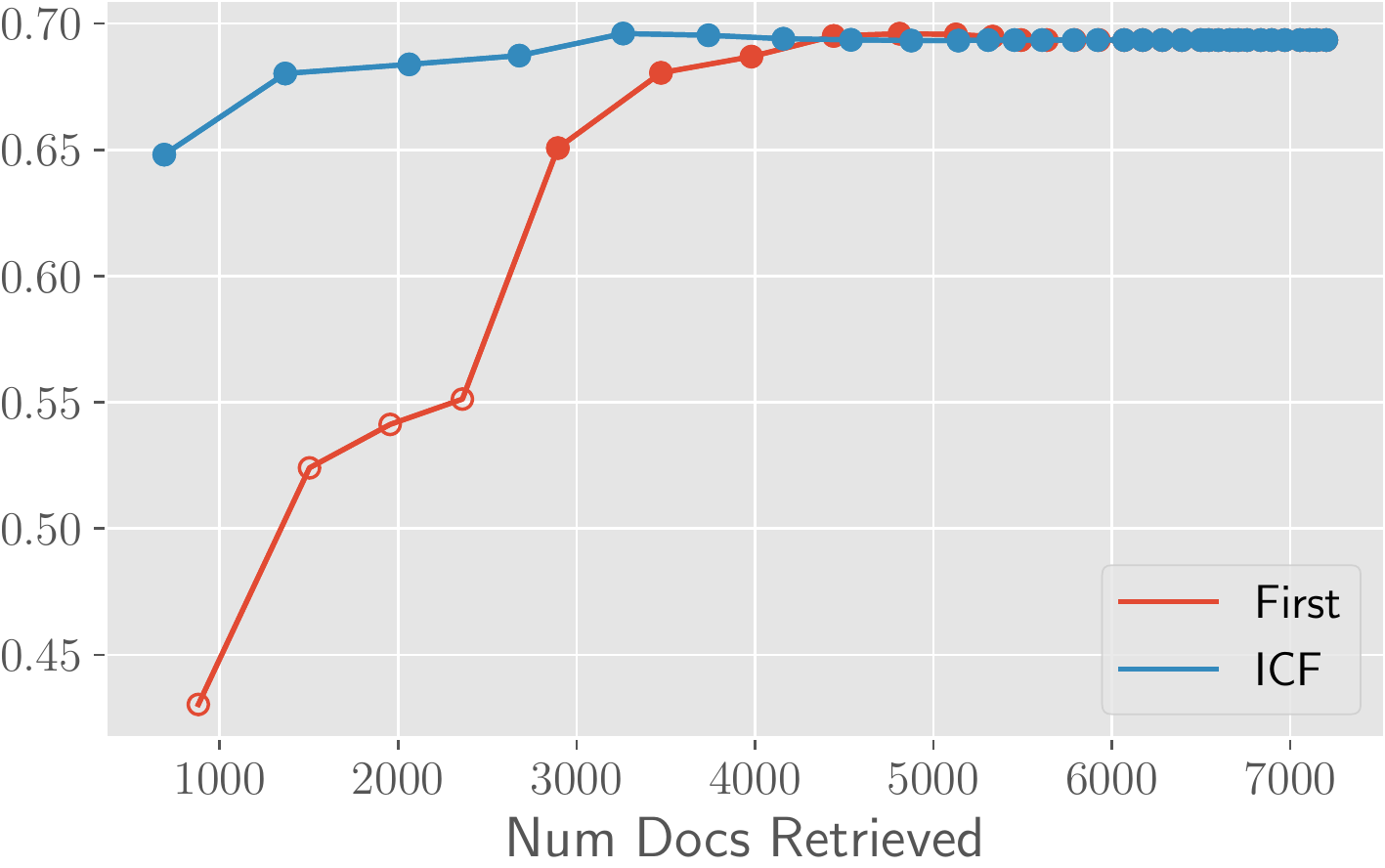}\vspace{-1.5mm}
    \caption{nDCG@10 on TREC2019}
    \end{subfigure}
    \begin{subfigure}[t]{.33\linewidth}
    \includegraphics[width=\figurewidth]{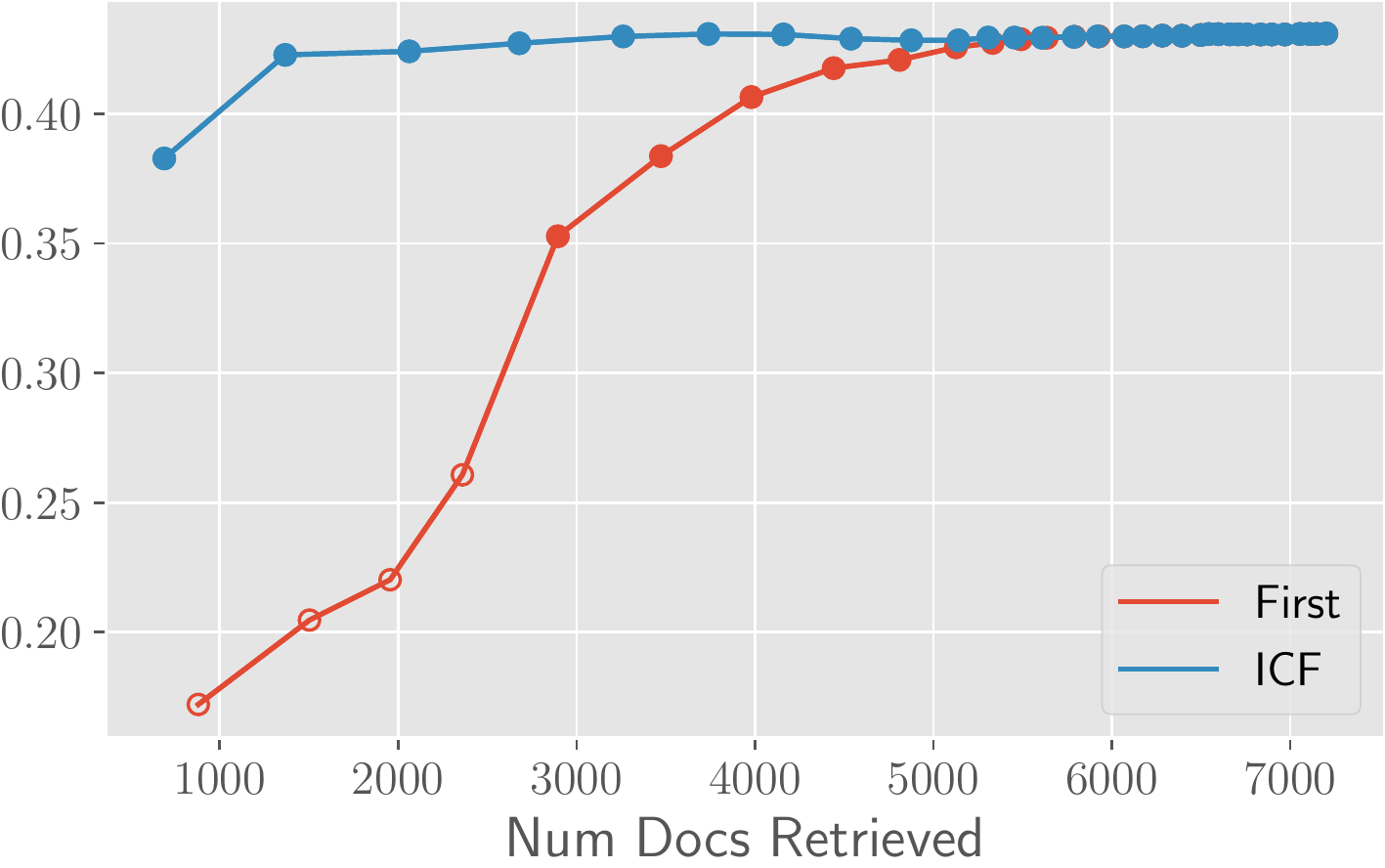}\vspace{-1.5mm}
    \caption{MAP on TREC2019}
    \end{subfigure}
    \begin{subfigure}[t]{.33\linewidth}
    \includegraphics[width=\figurewidth]{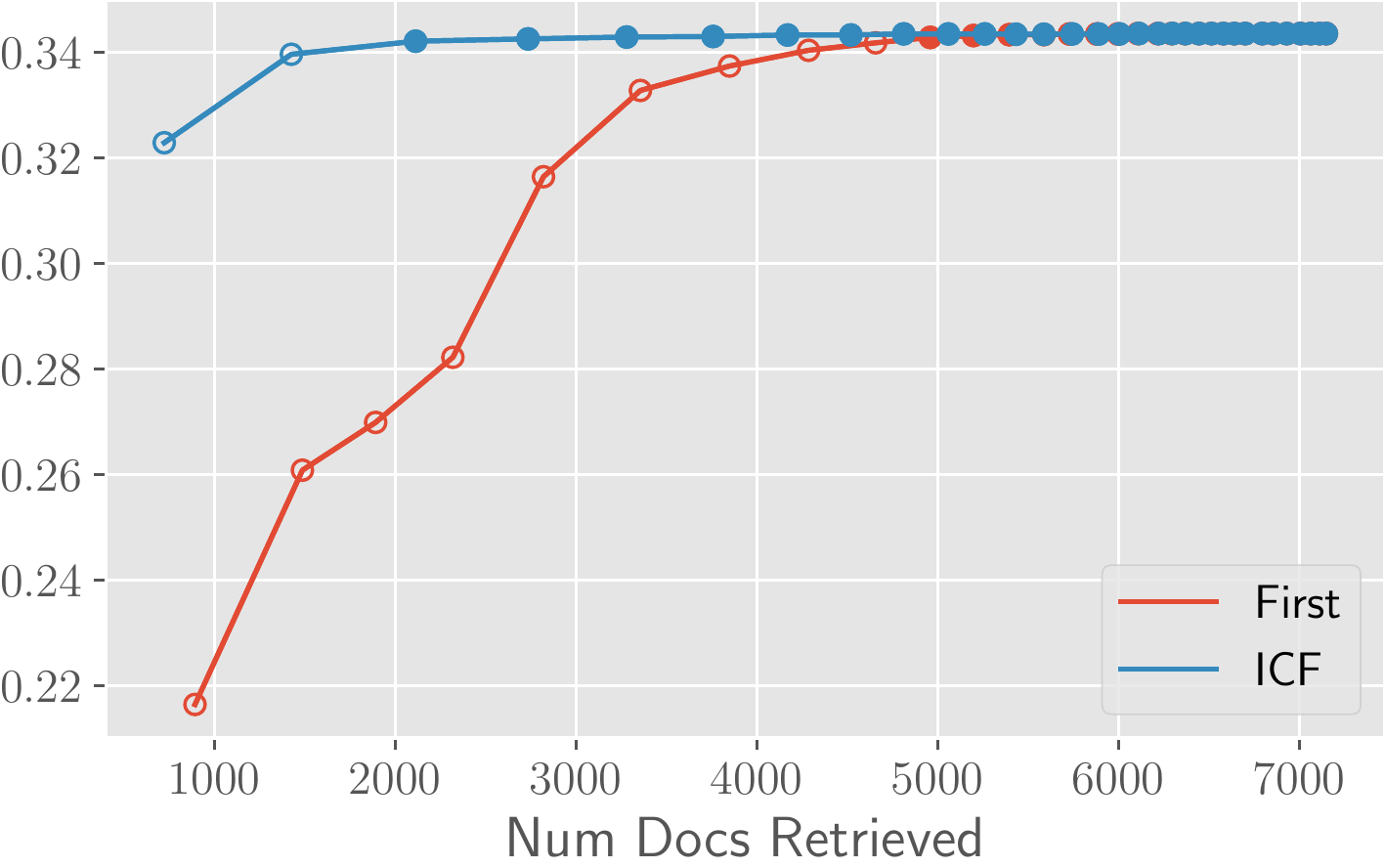}\vspace{-1.5mm}
    \caption{MRR@10 on MSMARCO Dev}
    \end{subfigure}\vspace{-.75\baselineskip}
    \caption{Effectiveness vs.\ number of retrieved documents as the number of query embeddings $p$ is varied for the baseline (First) and our proposed query embedding pruning (ICF). From left to right, each point on each curve is one additional query embedding, i.e. $p=1, 2, ..,  32$. \crc{Hollow} points denote a statistically significant difference in effectiveness compared to ColBERT $p=32, k'=1000$, according to a paired t-test with Bonferroni correction (p-value $< 0.05$).}
    \label{fig:results}
\end{figure*}
\looseness -1 Our experiments \crc{make use the MSMARCO passage ranking dataset and the PyTerrier IR experimentation platform~\cite{macdonald2020declarative,pyterrierCIKM}}. We use the ColBERT implementation provided by the authors\footnote{\url{https://github.com/stanford-futuredata/ColBERT}}, which we have extended\footnote{\crc{\url{https://github.com/terrierteam/pyterrier_colbert}}}. We follow~\cite{khattab2020colbert} for the settings of ColBERT; the resulting document embeddings index is 176 GB. The FAISS ANN index is trained on a randomly selected 5\% sample of the document embeddings. \craigf{The resulting FAISS index is 16 GB}. ANN search is performed on the 10 partitions most similar to the given input embeddings.
For evaluating effectiveness, we use the available querysets with relevance assessments: the official small version of the MSMARCO Dev set, consisting of 6,980 queries with on average 1.1 judgements per query, as well as the TREC 2019 queryset, which contains 43 queries with an average of 215.3 judgements per query\footnote{Additional experiments conducted on TREC 2020 confirmed our results.}. To measure effectiveness, we employ MRR@10 for the MSMARCO Dev queryset, and the MRR@10, nDCG@10 and MAP for the TREC queryset. We compare our results to the default setting of dense retrieval of ColBERT, using all $32$ query embeddings, and retrieving $k'=1000$ documents for each query embedding. 
In conducting our experiments addressing the efficiency, we determine the success of our query embeddings pruning strategy based on ICF\footnote{We do not report results using IDF since they match those obtained by using ICF.}, compared to the baseline approach, called First, by demonstrating that for a fixed number of embeddings, it attains effectiveness that is not significantly different from that of the default setting, while resulting in less documents being retrieved and re-scored by ColBERT's second-stage.
%
%For evaluating efficiency, we use a machine with 2x 16-core Intel Xeon @ 3.6GHz CPUs and 500GB of RAM, and making use of a single Titan RTX GPU card. \inote{All reported timings are the average of three repetitions.}

\pageenlarge{2} \looseness -1 The red and blue curves in Figure~\ref{fig:results} correspond to selecting query embeddings based on their order in the query (denoted First) and based on collection frequency (denoted ICF). In general, from each of the figures, we can see that first (red) curve always exhibits lower effectiveness with a similar number of documents retrieved. For instance, when using one query embedding, on MSMARCO Dev, First retrieves on average 824 documents and achieves an MRR@10 of 0.2165; in contrast, ICF retrieves less documents (724) and achieves a higher MRR@10 (0.3229). Similar trends are observed for nDCG@10 and MAP on the TREC 2019 query set, where effectiveness is always increased by using ICF compared to First, while retrieving a similar number of documents (689 for First vs. 880 for ICF). Moreover, ICF exhibits less statistically significant differences in effectiveness compared to First.

In general, the higher effectiveness of ICF over First is apparent for larger number of query embeddings, and effectiveness saturates at less query embeddings retrieving less documents: for instance, for nDCG@10 and MAP on TREC 2019 (Figures~\ref{fig:results}(a)~and~\ref{fig:results}(b)), at $p=2$, ICF reaches the same values as the full ColBERT retrieval with 32 query embeddings, but re-ranking 1367 documents only on average, while First needs at least $p=8$ query embeddings and 4441 documents on average. On MSMARCO Dev (Figures~\ref{fig:results}(c)), MRR@10 reaches the value of the full ColBERT retrieval with just the $p=3$ query embeddings with the highest ICF score.
\nicC{When reducing $p$ from 32 to 3, we decrease the number of documents retrieved by the first stage by 70\%, e.g., from {$\sim$}7000 to {$\sim$}2000. In terms of mean response time for the end-to-end to end system, this results in a reduction from 461.4ms to 173.7ms , i.e., a $2.65\times$ speedup.}

\looseness -1 \nicC{Note that the mean number of masked tokens in both query sets is 22.2, and, correspondingly, the average number of wordpieces per query is 9.8. In First, the query embeddings of the masked tokens appear last, hence their impact on the average effectiveness across all metrics is negligible. In ICF, \crc{all} special tokens \crc{(e.g.\ CLS \& MASK)}, appear last; in this case, the query embedding corresponding to the CLS token \crc{(on average the $\sim$10th query embedding under ICF)} does not contribute to the final effectiveness if the original query word tokens are used first.} Overall, we conclude from Figure~\ref{fig:results} that using the collection frequency of the query tokens can be used as indicator of the importance of the corresponding query embeddings, \nicC{and our proposed query embedding pruning strategy with 3 embeddings can obtain effectiveness results with no statistical significant differences w.r.t the original system using all query embeddings.} Indeed, the lower frequency tokens are more discriminative, and their query embeddings retrieve most of the relevant documents. This conclusion aligns well with the use of measures such as IDF in deciding on the most informative query terms for \taat\ dynamic pruning.

%% file: conclusions.tex
\pageenlarge{2}\section{Conclusions}\label{sec:conclusions}

In this paper, we identified efficiency challenges concerning the use of multiple embedding representations of queries and documents for dense retrieval. We proposed query embedding pruning, and demonstrated that a subset of the original query embeddings can be used for effective retrieval while reducing the number of document requiring to be exactly scored.
For example, when reducing the number of query embeddings used from 32 to 3, our approach results in no statistically significant differences in effectiveness, while reducing the number of documents retrieved and fully scored by 70\%. \nicC{In terms of mean response time for the end-to-end to end system, this results in a $2.65\times$ speedup.}
The results in this paper give rise to several possible direction of future work. The effectiveness of pruning suggests that adapting static pruning~\cite{10.1145/383952.383958} to work on embedding-based document representations before approximate nearest neighbour search may also have potential. Moreover, query embedding pruning can be applied selectively~\cite{tonellotto:2013}, with a different number of embeddings selected for different queries.

%\vspace{-0.1cm}
%\small
\section*{Acknowledgements}

Nicola Tonellotto was partially supported by the Italian Ministry of Education and Research (MIUR) in the framework of the CrossLab project (Departments of Excellence). Craig Macdonald acknowledges EPSRC grant EP/R018634/1: Closed-Loop Data Science for Complex, Computationally- \& Data-Intensive Analytics. 

%\craig{The results in this paper give rise to several possible direction of future work: Firstly, the effectiveness of dynamic pruning suggests that adapting static pruning~\cite{10.1145/383952.383958} to work on embeddings based document representations before approximate nearest neighbour search may also have potential; we also applied a fixed number of documents $k'$ to be retrieved for each query embedding, when a more advanced model may vary $k'$ for each query embedding. Finally, our \eaat\ query embedding learned models were simply based upon existing regression-tree models -- neural methods integrated into language model training may better recognise how many query/document embeddings are needed.}